\documentstyle{article}
\title{
The role of mathematics in contemporary theoretical physics
\thanks{Talk given at the 6th Philosophy-and-Physics-Workshop
``Epistemological Aspects of the Role of Mathematics in Physical
Science'', FEST, Heidelberg, Feb.\ 1993}
}

\author{Gernot M\"unster \\
        Institut f\"ur Theoretische Physik I \\
        Westf\"alische Wilhelms-Universit\"at \\
        Wilhelm-Klemm-Str.~9, D-4400 M\"unster, Germany}
\date{}

\begin{document}
\maketitle
\section{Introduction}
One of the subjects being discussed at this workshop is {\em the role of
mathematics in theoretical physics}.
As a physicist I have an idea of what theoretical physics is.
Looking at it from my own field of research, I may be allowed in the
present context to restrict myself to that part of theoretical physics,
which aims at a theoretical description of the basic elements of the
physical world: elementary particles and their interactions.

The physicists' view of mathematics, on the other hand, is always very
subjective and biased.
His own experience provides him only with a small window to contemporary
mathematics.
Therefore, when speaking about the role of mathematics in physics, I
will certainly raise more questions than give answers.
On the other hand, we are invited to have an open discussion, which
is not limited to safe statements.
So I feel encouraged to talk also about things which are outside my
competence, although it may be dangerous.
Concerning mathematics in general I have to rely, apart from my own
impression, on a random selection of authors, who have written about
developments, trends and concepts in past and present mathematics.
\section{Mathematics and Physics}
Let me start the discussion with the characteristic distinction between
mathematics and physics.
Physics is a science which aims at an understanding of the
regularities in the observed world.
Concerning mathematics, on the other hand, I may cite from I.\ Kant,
Kritik der reinen Vernunft, II.~Teil: ``Die phi\-lo\-so\-phi\-sche
Erkenntnis ist die Ver\-nunft\-er\-kennt\-nis aus Begriffen, die
mathematische aus der Konstruktion der Begriffe.''
Mathematics does not deal with phenomena and objects from the physical
world, but with intellectual constructs and their relations.
It is therefore counted as ``Geisteswissenschaft'' instead of
``Naturwissenschaft''.
Its objects are creations of the human mind.

However, we know that mathematics is the language of physics, as
has been expressed by Galilei, Kepler and many others later.
This leads us directly to the central topic of this workshop.
Mathematical concepts apply to physics.
How is this possible?
And how did it come about?
A look into history may be appropriate here.
\section{The relation between mathematics and
%\protect\newline
physics in history}
Here is one of the places where I walk on thin ice.
Being neither mathematician nor historian I have to rely on a small
selection of texts on the history of mathematics
\cite{Bell,Smith,Meschkowski,Bochner}.
The view presented will be oversimplified, but it is meant to reflect
general trends and features, even if it does not apply to every single
case.

The beginning of modern mathematics is often located in the
16th/17th century.
If we look into the hall of fame of modern mathematics we find among
others the names of Newton, Leibniz, Euler, Bernoulli, d'Alembert,
Lagrange, Legendre, Fermat, Laplace, Pascal, Hamilton and Gau\ss,
who are equally well known as great physicists.
This coincidence relates to the origin of many branches of modern
mathematics.
History reveals that many objects and concepts of modern mathematics
have their origin in physics.
Although counterexamples may certainly be found, this statement is
meant to be characteristic for the rise of modern mathematics.
In order to clarify it I shall give some examples.

i) Whereas Leibniz was more motivated by geometrical problems, Newton
came to invent calculus from the desire to have an adequate mathematical
instrument for the description of the motion of physical bodies.
Under the hands of Euler, Lagrange, Hamilton and others mechanics
developed into analytical mechanics.
Analytical mechanics in turn was the origin of the calculus of
variations, founded by Lagrange.

ii) The Hamilton-Jacobi formulation of mechanics was one of the main
origins of the theory of partial differential equations.
Considerations about partial differential equations led Sophus Lie to
the concept of continuous groups.

iii) Modern differential geometry has different origins.
One of them is related to Gau\ss' duty to perform geodesic
measurements in the county of Hannover, and to his studies of planetary
motion.
In this context he developed the theory of curves and surfaces of the
17th and 18th century into classical differential geometry.
On the other hand there is the foundation of tensor calculus by Ricci
and Levi-Civita, which was partly motivated by the use of generalized
coordinates in analytical mechanics.
Elasticity theory also contributed here.

iv) The investigation of the nature of orbits of classical dynamical
systems by Poincar\'e gave rise to the foundation of modern topology by
him.

v) In the area of functional analysis we may also observe various
physical roots.
The theory of differential equations, which was advanced by Euler and
others, was motivated to a large extent by physical problems.
A particular differential equation, the heat equation, led to the theory
of Fourier series, which in turn is one of the starting points of
Hilbert space theory.
Another root of Hilbert spaces and Banach spaces is the theory of
integral equations (Fredholm, Hilbert, Schmidt), which emerged from
physical and technical problems.

In the early phase of modern mathematics, to which the examples refer,
there was a close relation between mathematics and physics, as the above
mentioned personal union of great mathematicians and physicists shows.
Mathematics developed in direct contact with theoretical physics of that
time.

This started to become different in the following historical phase.
The characteristic element of the subsequent development of mathematics
is {\em abstraction}.
With the beginning of the 19th century mathematics became more and more
independent of mechanics and then physics.
Mathematics began to undertake abstractions as a major element of its
development \cite{Bochner}.
This change is not meant to have occurred to mathematics in total, but
to different branches at different times according to the state of their
historical development.
Analytic geometry, via linear algebra evolved into spectral theory;
the theory of algebraic equations gave rise to abstract group theory and
modern general algebra; algebraic geometry, topological K-theory,
commutative algebra and other abstract fields of mathematics arose,
to give a few keywords only.

This process of abstraction is a natural and essential element of the
way of mathematics.
Y.~Manin writes \cite{Manin}: ``\ldots mathematics associates to some
important physical abstractions (models) its own mental constructions,
which are far removed from the direct impressions of experience and
physical ex\-pe\-ri\-ment.''
The role of abstraction became particularly evident in our century.
According to Dieudonn\'e \cite{Dieudonne} the main concern of
mathematics since 1920 is the study of structures rather than objects.
This shows up in the importance of concepts like invariants and
categories.
The invention of structures became a main part of mathematical
activities.
Many examples supporting this view can be found in Dieudonn\'e's article
\cite{Dieudonne}.

The tendency towards abstraction in mathematics of course had its effect
on the relation between mathematics and physics.
Whereas Y.\ Manin summarizes it by the statement ``in this century
[mathematics and physics] have both engaged in introspection and
internal development'', F.J.\ Dyson does not hesitate to speak of the
divorce of mathematics and physics at the turn of the 19th to the 20th
century \cite{Dyson1}.

What is the further evolution of the relation between these two
sciences?
A new convergence could be expected, if R.\ Courant's \cite{Courant}
characterization of the general development of a mathematical discipline
is correct:
``Generally speaking, such a development will start from the `concrete'
ground, then discard ballast by abstraction and rise to the lofty layers
of thin air where navigation and observation are easy; after this flight
comes the crucial test of landing and reaching specific goals in the
newly surveyed low plains of individual `reality'.''

Before I turn to this question directly, I would like to make a
digression to Wigner's thesis, which is related to it.
\section{The unreasonable effectiveness of mathematics in the natural
sciences}
This is the title of a much cited article by E.P.\ Wigner \cite{Wigner}.
In it Wigner points out the physicists observation that mathematical
concepts turn up in entirely unexpected connections and then often
permit an unexpectedly close and accurate description of phenomena.
Here he refers to advanced mathematical concepts, which are abstract and
are not directly suggested by the actual world.
Wigner considers them not to be simple, but being made by the
mathematicians in order to allow ``brilliant manipulations''.
What he calls the unreasonable effectiveness of such concepts means
that they apply very efficiently in e.g.\ physics.

{}From the various examples given in the article \cite{Wigner} I would
like to mention only the case of quantum mechanics.
The theory of Hilbert space and linear operators is an abstraction,
which has its roots in places which have nothing to do with quantum
phenomena.
A priori there is no reason to expect that interference patterns
produced on a screen by streams of matter have anything to do with
vectors in a Hilbert space.
Nevertheless it turned out that quantum phenomena are most adequately
described in terms of the mathematics of Hilbert spaces.
Moreover this mathematical structure did not remain restricted to the
physics of atomic spectra, where it was introduced, but turned out to be
of such a general validity that it governs the physics of molecules,
nuclei, solid bodies, stars etc..

The last mentioned aspect is another related point made by Wigner: the
simplicity of the basic laws goes along with a large range of
applicability.
In the same spirit H.\ Hertz wrote about Maxwell's equations (cited in
\cite{Dyson2}):
``One cannot escape the feeling that these mathematical formulae have an
independent existence and an intelligence of their own, that they are
wiser than we are, wiser even than their discoverers, that we get more
out of them than was originally put into them'', and Dyson writes
\cite{Dyson2}: ``As often happens in physics, a theory that had been
based on some general mathematical arguments combined with few
experimental facts turned out to predict innumerable further
experimental results with unfailing and uncanny accuracy.''

According to Wigner the enormous usefulness of mathematics is
unexpected inasmuch as there is no rational explanation for it.
{}From an epistemological point of view the question for a rational
explanation is of central importance.
One possibility is suggested by history as discussed above: the
mathematical concepts under consideration are just made such as to fit
physics.
E.\ Mach has introduced the term ``Denk\"okonomie'' to describe this
relation \cite{Mach}.
This means that theories in the natural sciences are made such as to
yield a highly economical description of phenomena.
In some cases this may be true.
But there are arguments against such an explanation: 1.) in the examples
discussed by Wigner and Dyson the mathematics was prefabricated for
different purposes, 2.) an economy of description of present knowledge
cannot be expected to yield such a big success for future predictions,
3.) the really surprising fact is the existence of such highly
economical descriptions.
Newton's law of gravitation for example was derived from rather crude
observations, but nowadays we know that it holds extremely accurately
and universally.

Another attempt to explain the effectiveness of mathematics, expressed
by N.\ Bohr, points out that mathematics, as a science of structures,
provides us in principle with all possible structures.
Therefore it may not be surprising that the structures adequate for
physics are present in mathematics.
This argument does, however, not appear convincing to me.
Mathematics provides us potentially with all possible structures, but it
does not do so actually.
The structures studied by mathematicians are certainly only a tiny
fraction of all possible ones, but the ones most useful to physicists
are often just among the already existing, according to Wigner.

This leads me to a point where I would like to depart a little bit from
Wigner's arguments.
Wigner considers the abstract mathematical structures, which later turn
out to be useful in physics, as not being simple but to be sophisticated
constructs, which are chosen by the mathematicians because they allow to
exercise nice mental acrobatics.
But on the other hand, considering the process of abstraction and the
emergence of new concepts in mathematics, it appears that these new
concepts are often directly suggested naturally by the existing
formalism.
The existing formalism often intrinsically points out the ways to
surmount it.
This can be seen clearly in the cases of extensions of a number
system.
The ideal numbers, for example, were introduced by Kummer in order to
uphold the unique decomposition into primes in cyclotomic number fields.
Abstraction then led to the concept of ideals, introduced by Kronecker,
which is a central element in abstract algebra.

In this sense extensions of the mathematical formalism are in many cases
natural and far from arbitrary.
\section{Recent situation}
Wigner's thesis concerning the relation between mathematics and physics
is also predictive.
If we subscribe to it we should expect to find more examples of
mathematical concepts with unexpected applications in physics since the
1960's.
In this section I would like to take a look on the recent history of the
two fields in view of their relation.
There are two aspects which appear to be important for the recent
developments: unification, on the one hand, and renewed relation between
mathematics and physics on the other hand.

{\it Unification.}
In physics one observes an increasing tendency towards unification
concerning both its subject and its methods.
The trend to unification in the description of physical phenomena is
well known.
Quantum field theory established itself as the general universal
framework for the description of matter and forces.
Moreover the known particles and interactions apart from gravity could
be fitted into a unified theory, the so called Standard Model, which
contains Quantum Chromodynamics and the Glashow-Weinberg-Salam-Theory.
The unifying principle underlying the interactions, including gravity,
is the principle of local gauge invariance.
Respecting this principle attempts towards even more unified theories,
such as Grand Unified Theories and supersymmetric theories, are being
made.

Concerning the methods of theoretical physics a certain amount of
unification can also be perceived.
In particular statistical physics and elementary particle physics
experienced a big methodological overlap.
I shall only mention the keywords functional integrals, renormalization
group and scaling in this connection.

In mathematics Manin \cite{Manin} also sees unifying tendencies.
The theory of operator algebras, group theory, differential geometry,
topology, and algebraic geometry are some of the areas whose
interrelation and mutual influence seem to grow and lead to more
unifications.

On the other hand there also appears a renewed interest in specific
concrete objects.
The study of imaginary quadratic number fields, L-series, transcendental
numbers are just a few examples.

The driving forces for unification are of course different in
mathematics and in physics.
In mathematics it is the desire to find unifying concepts and
structures, whereas in physics one reaches for a unified description of
fundamental phenomena in terms of theoretical models.

{\it Relation between mathematics and physics.}
As has been mentioned above, theories with a local gauge invariance have
taken a prominent role in theoretical physics.
Looking into the literature of the past twenty years one finds that
there are various mathematical fields and concepts relevant for the
treatment of the physics of gauge fields.
Differential geometry and fibre bund\-les turned out to represent the
proper mathematical framework for gauge field theories; topology and
characteristic classes played an important role in connection with
instantons, and the physics of anomalies found its adequate description
in terms of cohomology theory, index theorems and related items, to
mention some examples.

This is clearly another instance of Wigner's observation.
Differential geometry, in particular principal fibre bundles,
Cartan-Ehresmann-connections etc.\ were developed on the basis of
mathematical reasoning which had nothing to do with non-Abelian gauge
theories, which were only introduced in theoretical physics in the late
1950's.
Nevertheless they appear to yield the mathematical structures which fit
perfectly well the needs of gauge theories.

Conversely, there is an interesting feedback from physics to mathematics
in recent history.
Considerations about instantons, which are solutions of the gauge field
equations, have led to astonishing new results about the geometry of
four-dimensional manifolds.
Conformally invariant quantum field theories and field theories with
Chern-Simons action have brought out new fruitful approaches to knot
theory and to the topology of three-dimensional manifolds.
Even more fascinating in this respect is the case of string theory,
which employs as different topics as Riemann surfaces, algebraic
geometry, lattices, infinite-dimensional Lie algebras, modular forms,
number theory and others, and has revealed new connections between these
fields, one example of which is known under mathematicians as
``monstrous moonshine''.

In accordance with the view presented above Manin \cite{Manin} finds
that the present relation between mathematics and physics is
characterized by an increasing overlap between them, and he observes a
growing willingness to learn from each other and to transfer tools and
techniques.
With respect to Wigner's thesis I think that the recent developments are
in support of it in even more unexpected circumstances.

In closing I would like to cite R.\ Courant again \cite{Courant}:
``That mathematics, an emanation of the human mind, should serve so
effective for the description and understanding of the physical world is
a challenging fact that has rightly attracted the concern of
philosophers.''
\end{document}